\documentclass[aps,prl,preprint,amsmath,amssymb]{revtex4}
\usepackage{graphicx}
\usepackage{dcolumn}
\usepackage{bm}

\begin{document}

\title{Entropy of jammed matter}

\author{Christopher Briscoe, Chaoming Song, Ping Wang, Hern\'an
A. Makse}

\affiliation {Levich Institute and Physics Department, City
College of New York, New York, NY 10031, US}

\date{\today }

\begin{abstract}
We investigate the nature of randomness in disordered packings of
frictional spheres. We calculate the entropy of 3D packings through
the force and volume ensemble of jammed matter, a mesoscopic ensemble
and numerical simulations using volume fluctuation analysis and graph
theoretical methods.  Equations of state are obtained relating
entropy, volume fraction and compactivity characterizing the different
states of jammed matter.  At the mesoscopic level the entropy vanishes
at random close packing, while the microscopic states contribute to a
finite entropy. The entropy of the jammed system reveals that the
random loose packings are more disordered than random close packings,
allowing for an unambiguous interpretation of both limits.
\end{abstract}

\maketitle

Filling containers with balls is one of the oldest mathematical
physical puzzles known to scientists \cite{bernal}. Apart from its
mathematical significance, this problem has found applications in
modern topics, such as jamming in granular media, colloids, and
the structure of liquids and the glass transition \cite{coniglio}.
Despite vast progress made in developing a statistical mechanics
for such systems \cite{sirsam}, basic questions remain unanswered,
including \cite{bernal,berryman,salvatore2}: What is a jammed
state and how to characterize its state of randomness?

\begin{figure*}[t]
\centering { \hbox { (a)
\resizebox{4cm}{!}{\includegraphics{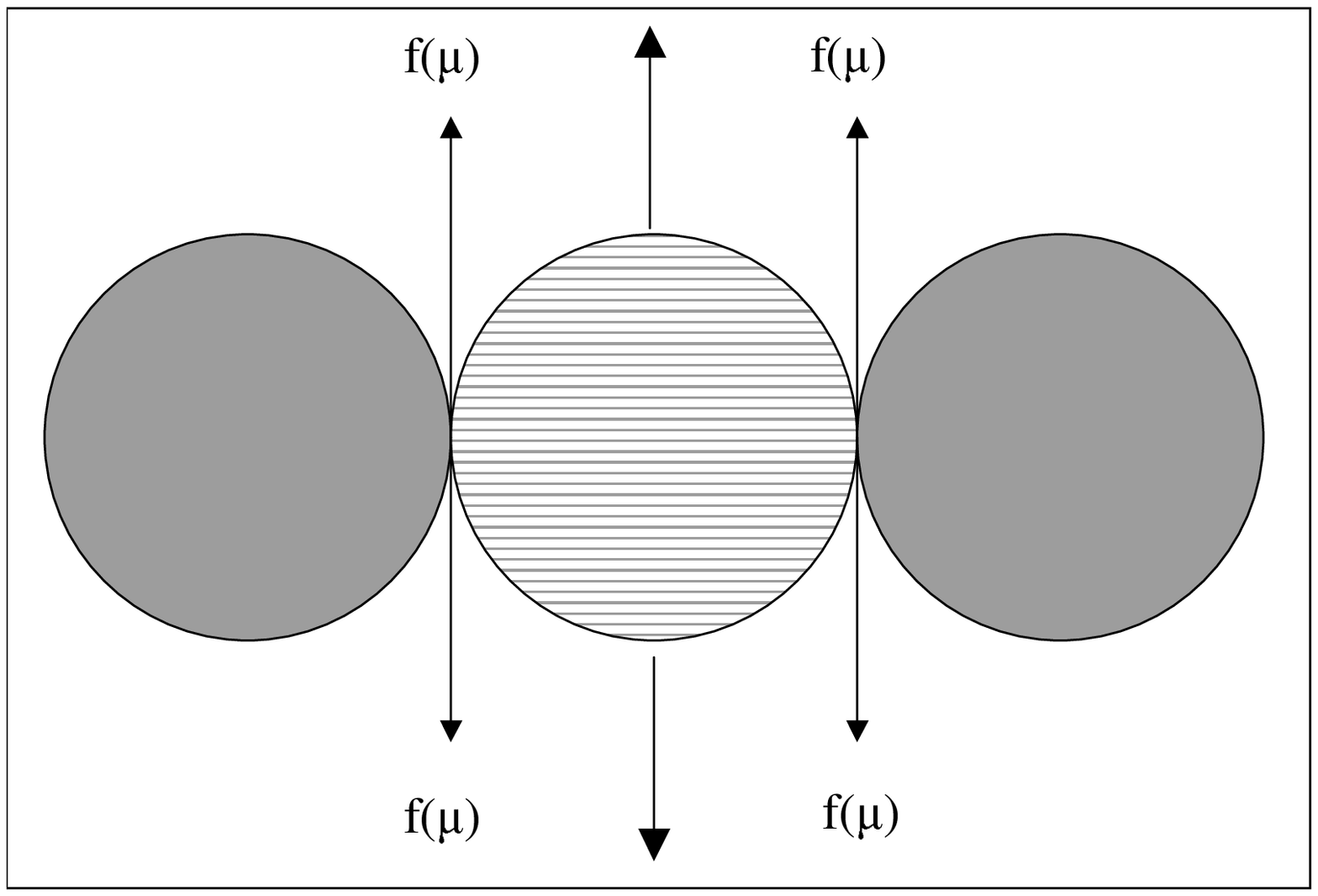}}(b)
\resizebox{5cm}{!}{\includegraphics{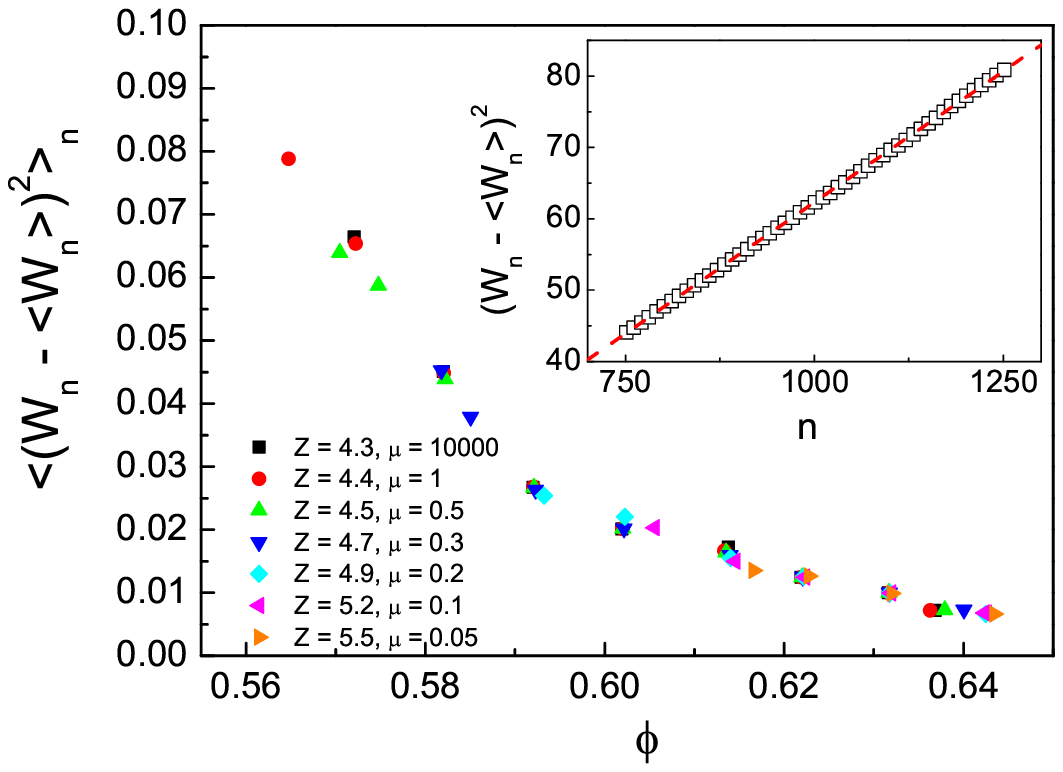}} (c)
\resizebox{5cm}{!}{\includegraphics{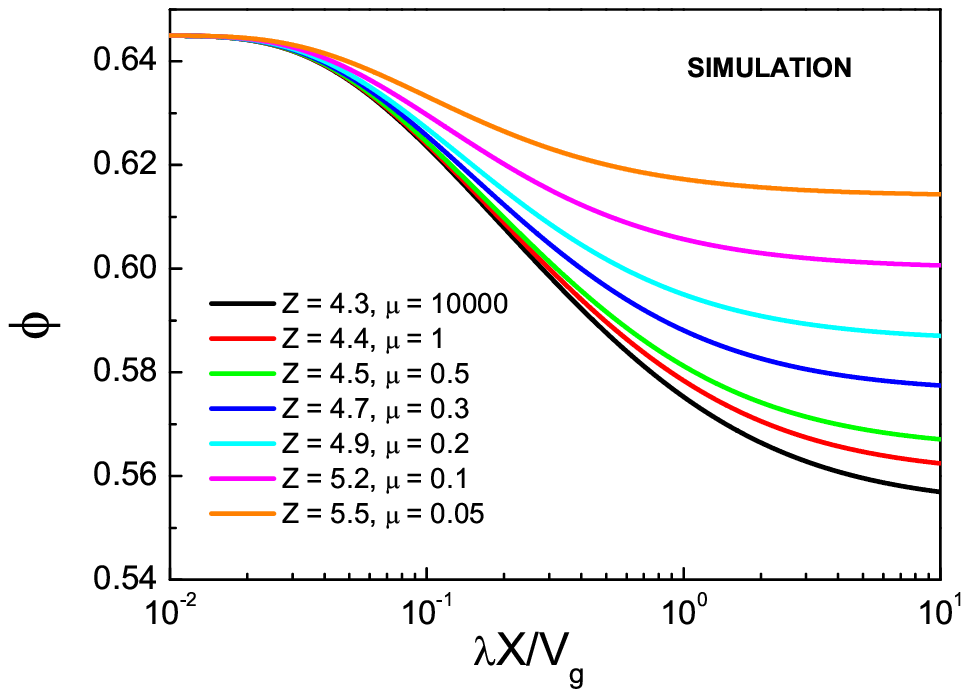}}} \hbox {(d)
\resizebox{5cm}{!}{\includegraphics{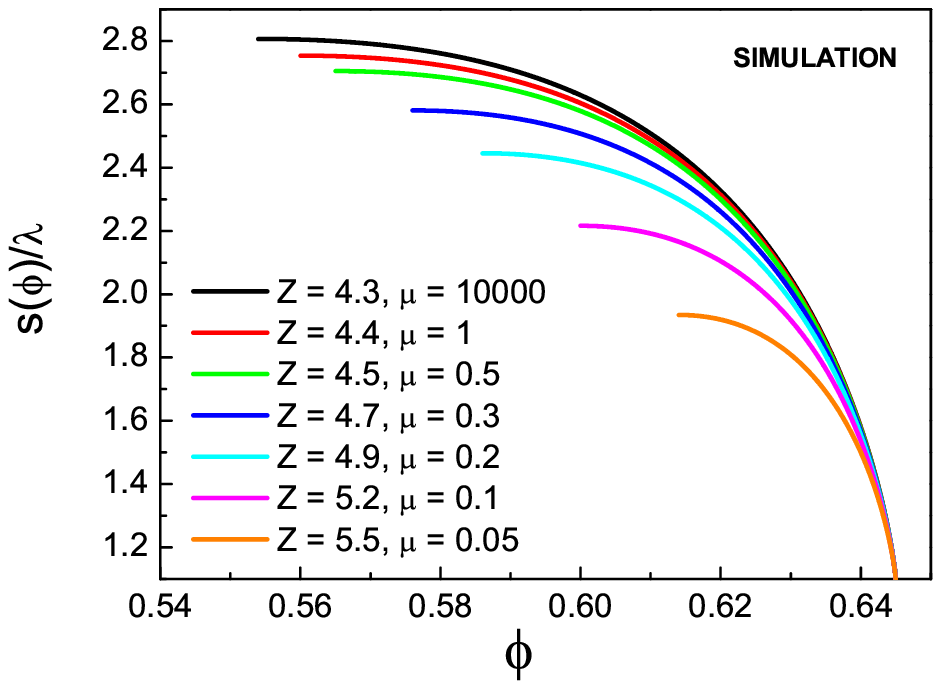}} (e)
\resizebox{5cm}{!}{\includegraphics{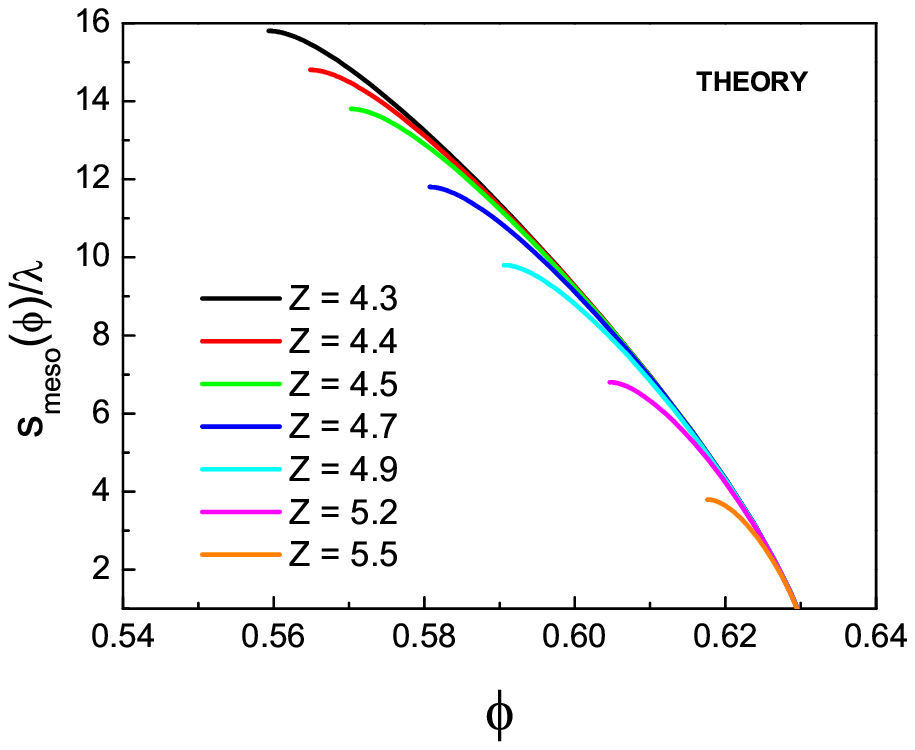}} (f)
\resizebox{5cm}{!}{\includegraphics{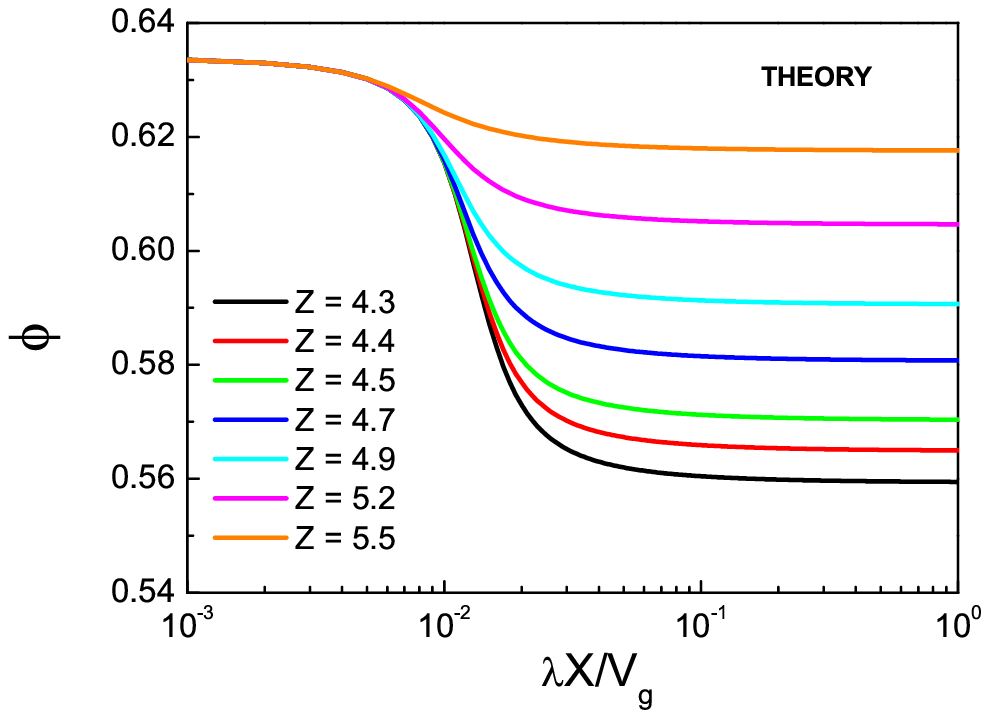}} } }
\caption{Numerical and theoretical results. (a) A disk in 2d under
mechanical equilibrium by two nearest neighbor contacts. The disk
is not jammed under a normal force interaction, but it jams when
tangential forces are present.  (b) $\langle \Delta {\cal
W}_n^2\rangle_n $ versus $\phi$. Inset shows the extensivity of
$\langle \Delta {\cal W}_n^2\rangle $ versus $n$. (c) $\phi$
versus $X$ from the integration of (b). (d) Entropy versus $\phi$
from the integration of (c).  Entropy at RCP achieves a value of
1.1$\lambda$ as calculated by the Shannon entropy at RCP. (e)
Prediction of the mesoscopic theory for $s_{\rm
meso}(\phi)/\lambda+1.1$. (f) Prediction of the mesoscopic theory
for $\phi(X)$.} \label{numerics}
\end{figure*}

In an attempt to define the jammed states in a rigorous way,
Torquato and coworkers have proposed three categories of jamming
\cite{salvatore1}: locally, collectively and strictly jammed. This
problem is intimately related to the existence of well-defined
upper and lower limits in the density of disordered packings;
random close packing (RCP) and random loose packing (RLP)
\cite{bernal}; a longstanding open question in the field.  A
definition of RCP requires proper definitions of {\it jammed
states} and the concept of {\it randomness} \cite{salvatore2}.
These previous definitions of jamming are based purely on geometrical
considerations, sufficient to describe frictionless grains
\cite{salvatore1} but not for granular materials where friction
dominates \cite{kertesz}. Figure \ref{numerics}a illustrates the
point: a frictionless hard sphere system is not locally jammed if
only normal forces are considered, since the ball can freely move
in the vertical direction. The same geometrical configuration is
locally jammed if friction is allowed between the particles,
revealing the importance of forces in the definition of jamming
for frictional particles.

The goal of the present paper is to characterize jamming and the
degree of randomness for frictional hard spheres and obtain equations
of state. Our framework is rooted in statistical mechanics and
considers calculation of the entropy at the volume-force (V-F)
ensemble level \cite{sirsam} through force and torque balance
conditions.  The simulations and theory developed allow understanding
of the qualitative behavior of equations of state relating volume
fraction, entropy and compactivity, and illuminate the nature of RCP
and RLP. We follow three complementary approaches to characterize the
entropy of jammed matter.  (a) Computer studies: We first investigate
frictional packings of spheres at the jamming transition
numerically. We compute the equations of state, entropy and
compactivity, as a function of volume fraction, ranging from RLP to
RCP.  The entropy is calculated by two numerical methods, direct
analysis of volume fluctuations via a fluctuation-dissipation theorem
and graph theoretical methods. Simulations show that random loose
packings are more disordered and have higher compactivity than random
close packings. In order to rationalize the simulation results we
follow two theoretical approaches: (b) We develop the concept of
randomness in the V-F ensemble following the Gibbs distribution
\cite{sirsam}, which is different from the measurement of randomness
of a single packing in terms of an ensemble of order parameters
\cite{salvatore2}. Since analytical expressions of the entropy are
difficult to obtain in this ensemble we develop, (c), A mesoscopic
ensemble approximation \cite{jamming2} based on mechanical equilibrium
imposing an average coordination number, $Z$, larger or equal than the
minimum isostatic coordination as conjectured by Alexander
\cite{alexander} (see also
\cite{stealing,makse,ohern,kertesz,silbert}). Calculations are done
under the mesoscopic approximations of \cite{jamming2} giving rise to
a mesoscopic configurational entropy, achieving a minimal value at RCP
and maximal value at RLP. The results characterize the disorder of RCP
and RLP at the mesoscopic level in general agreement with the
simulations. They also suggest that the configurational entropy
requires augmentation to include the entropy of the microscopic states
neglected at the mesoscopic level.

{\it Numerical simulations.---} We investigate computer generated
packings of 10,000 spherical equal-size particles of 100$\mu$m
diameter interacting via Hertz (normal) and Mindlin (tangential)
contact forces (with shear modulus 29GPa and Poisson's ratio 0.2)
with Coulomb friction, $\mu$, using methods previously developed
in \cite{makse,jamming2}.  Packings characterized by different
$\mu$ are generated  by compressing a gas of particles from an
initial (unjammed) density, $\phi_i$, with a compression rate,
$\Gamma$, until a final density, $\phi$, at the jamming
transition.

The friction $\mu$ ranges from 0 to $\infty$ producing packings
with coordination number varying from $Z\approx 6$ to $Z\approx
4$, respectively.  Thus, the preparation protocol produces
packings with densities parameterized by
$\phi(\mu,\Gamma,\phi_i)$. In general, the lower $\Gamma$ the
smaller the obtained $\phi$, while the larger $\phi_i$ the larger
$\phi$ of the packing.  We also find that under our numerical
protocol there exits a common function $Z(\mu)$ over the different
$\Gamma$ and $\phi_i$ (see \cite{jamming2}).  For $\mu\to\infty$,
$\phi$ ranges from the RLP limit $\phi_{\rm RLP}\approx 0.55$
obtained when $\Gamma\to 0$ and $\phi_i<0.55$ to the RCP limit
$\phi_{\rm RCP}\approx 0.64$ obtained for larger $\Gamma$ and
$\phi_i\to 0.64$ (note that merely changing $\Gamma$ is not
sufficient to allow $\phi_{\rm RCP}$ to approach $0.64$, as
discussed in \cite{jamming2}, for $\mu \rightarrow \infty$). For
$\mu=0$, the density is approximately $\phi\approx\phi_{\rm RCP}$
for any $(\Gamma,\phi_i)$.  For intermediate $\mu$, the packings
follow the phase diagram as obtained in \cite{jamming2}.

The calculation of the entropy as a function of volume fraction is
realized by using volume fluctuation analysis through a
Fluctuation- Dissipation relation \cite{chicago2} complemented
with graph theoretical methods \cite{shannon,vink}.  We first
define the Voronoi cell associated with each particle $i$ and
calculate its Voronoi volume ${\cal W}_i$. We then perform
statistical analysis of the volume fluctuations by considering a
cluster of $n$ contacting particles with volume ${\cal W}_n =
\sum_i^n {\cal W}_i $.  We calculate the average volume, $\langle
{\cal W}_n\rangle$ and fluctuations $\langle ({\cal W}_n - \langle
{\cal W}_n\rangle )^2 \rangle$, where $\langle \cdot \rangle$ is
an average over many $n$-clusters. We find that for sufficiently
large $n\approx 1000$, contrasting the results of \cite{dauchot},
the fluctuations scale with $n$ and therefore are extensive and
well-defined (see inset of Fig. \ref{numerics}b).

From the large $n$ behavior we extract the fluctuations, plotted
in Fig \ref{numerics}b, versus $\phi$ for every packing studied.
The compactivity of the packing, $X$, is obtained via the
integration of the fluctuation relation $\langle ({\cal W}_n -
\langle {\cal W}_n\rangle )^2 \rangle = \lambda X^2 d\langle {\cal
W}_n\rangle / dX$ as $X^{-1} = \lambda \int_{\phi(X)}^{\phi_{\rm
RLP}} d\langle {\cal W}_n\rangle/ \langle ({\cal W}_n - \langle
{\cal W}_n\rangle )^2 \rangle$, where we use that $\phi(X\to
\infty) \to \phi_{\rm RLP}$ \cite{jamming2}, and $\lambda$ is the
analogue of the Boltzmann constant. Since Voronoi volumes are
additive, $\langle {\cal W}_n\rangle = \langle {\cal W}\rangle =
NV_g/\phi$, where $V_g$ is the volume of the grain. Therefore, the
above integration is rewritten as:

\begin{equation}
(X/V_g)^{-1} = \lambda \int_{\phi_{\rm RLP}}^{\phi(X)} d\phi/
(\phi^2\langle ({\cal W}_n - \langle {\cal W}_n\rangle )^2_n
\rangle), \label{compactivity}
\end{equation}
where $\langle ({\cal W}_n - \langle {\cal W}_n\rangle )^2_n$ is
the fluctuation density. We may then utilize the fluctuation
density as a function of $\phi$, shown in Fig. \ref{numerics}b,
and integrate along a line of constant $Z(\mu)$.  We note that
while the fluctuation densities for all $Z(\mu)$ in this study
collapse onto a single curve, illustrated in Fig \ref{numerics}b,
the limit of integration, $\phi_{\rm RLP}$ in Eq.
(\ref{compactivity}), changes as discussed in the phase diagram of
\cite{jamming2}, increasing as $\mu$ decreases.

The equation of state, $\phi(X)$, is plotted in Fig.  \ref{numerics}c
for different values of the average coordination number of the
packings, $Z(\mu)$, revealing that as we approach $\phi_{\rm
  RCP}\approx 0.64$, $X \to 0$, regardless of the value of
$\mu$. Further, $X \to \infty$ as we approach $\phi_{\rm RLP}$,
with the smallest volume fraction of the RLP appearing for $\mu\to
\infty$ and $Z\approx 4$, with $\phi_{\rm RLP}\approx 0.55$.

The entropy, $S$, and its density, $s=S/N$, are obtained by
integrating $(X/V_g)^{-1}= -\phi^2 \partial s/\partial\phi$, as:

\begin{equation}
s(\phi) - s(\phi_{\rm RCP}) = \lambda \int^{\phi_{\rm RCP}}_{\phi}
d\phi/[(X(\phi)/V_g)\phi^2]. \label{entropy}
\end{equation}

This analysis provides the entropy up to a constant of integration
$s(\phi_{\rm RCP})$. To obtain the entropy of RCP we use an
independent method based on information theory \cite{shannon,vink}
to obtain another estimation of the entropy. We use the Voronoi
cell and Delaunay triangulation for each particle to define a
Voronoi network. We construct a graph as a cluster of $n$
contacting particles which, by means of graph automorphism, can be
transformed into a standard form or "class" $i$ of topologically
equivalent graphs considered as a state with an occurrence $p(i)$.
In practice, we determine $p(i)$ by extracting a large number $m$
of clusters of size $n$ from the system and count the number of
times, $f_i$, a cluster $i$ is observed, such that: $p(i)=f_i/m$.
The Shannon entropy of a clusters of size $n$ is $H(n)=-\lambda
\sum p(i) \ln p(i)$, and the entropy density is $s=\lim_{n \to
\infty} [H(n+1)-H(n)]$, converging so rapidly that even moderate
values of $n$ are enough to obtain a sufficient approximation of
$s$ \cite{vink}.  The Shannon entropy density provides an
estimation of the entropy for the RCP state, $s(\phi_{\rm
RCP})\approx 1.1 \lambda$, serving as the constant of integration
for the entropy density as realized by volume fluctuations.  The
resulting entropy density is plotted in Fig. \ref{numerics}d
versus $\phi$ for different $Z(\mu)$.

When comparing all packings with different $Z(\mu)$ and $\phi$,
the maximum entropy is at the minimum volume fraction of RLP
$\phi_{\rm RLP}\approx 0.55$ when $X\to\infty$ at infinite
friction. The minimum entropy is found for the RCP state at
$\phi_{\rm RCP}\approx 0.64$ for $X\to 0$, for all the values of
friction, indicating the degeneracy of the RCP state, contrary to
the common belief that the RCP limit corresponds to a state with
the highest number of configurations and therefore the highest
entropy.

{\it Statistical mechanics of frictional hard spheres.---} Next,
we use the ensemble of jammed matter to rationalize the obtained
equations of state \cite{sirsam}.  Experiments of shaken grains,
fluidized beds and oscillatory compression of grains
\cite{chicago2} indicate that granular materials show reversible
behavior, and the analogue of the conserved energy, $E$, in
thermal systems is the volume $V=N V_g/\phi$, for a system with
$N$ at positions $\vec{r}_i$.  Thus, the number of configurations,
$\Omega$, and the entropy in the microcanonical ensemble of jammed
hard spheres is defined as \cite{sirsam}

\begin{equation}
\Omega(V) = e^{S(V)/\lambda} = \int \delta (V-{\cal W}(\vec{r}_i))
\Theta_{\rm jam}(\vec{r}_i)  {\cal D}\vec{r}_i \label{omega}
\end{equation}

Just as $\partial E/\partial S = T$ is the temperature in
equilibrium system, the ``temperature'' in granular matter is
$X=\partial V/\partial S$.  Here $\Theta_{\rm jam}(\vec{r}_i)$ is
a constraint function restricting the integral to the ensemble of
jammed states, ${\cal W}(\vec{r}_i)$ is the volume function
associated with each particle taking the role of the Hamiltonian.
The crux of the matter is then to properly define $\Theta_{\rm
jam}$ and $\cal W$ to calculate the entropy and volume in the
ensemble of jammed matter.

{\it Volume and force V-F-ensemble.---} A minimum requirement of
$\Theta_{\rm jam}(\vec{r}_i)$ is to ensure touching grains, and
obedience to Newton's force and torque laws. As in the numerical
simulations, the volume function, ${\cal W}(\vec r_i)$, is taken
as the volume of the Voronoi cell associated with each particle at
position $\vec r_i$, for which an analytical form has been
obtained in \cite{jamming2}.  Following Eq. (\ref{omega}), the
entropy in the V-F-ensemble of frictional hard spheres takes the
form:
\begin{equation}
\begin{split}
\Omega_{\rm VF}& = e^{S(V)/\lambda} = \int
\delta\big(V- {\cal W}(\vec{r}_i)\big)  \times \\
\prod_i \Big\{ \delta \big(\sum_{j \neq i} \vec{f}_{ij} \big) &
\delta\big(\sum_{j \neq i} \vec{f}_{ij} \times \vec{r}_{ij}\big)
\,\, \delta(\vec f_{ij} - \vec f_{ji}) \times \\
\prod_{j \neq i} \Big[\Theta(\mu f_{ij}^{\ N} - f_{ij}^{\ T})  &
\delta \big([(\vec r_{ij})^2-1]\big) {\cal D} f_{ij} \Big] {\cal
D} r_i \Big\},
\end{split}
\label{exact}
\end{equation}
where $\vec r_{ij} \equiv \vec r_i - \vec r_j$, the normal
inter-particle force is $f_{ij}^{\ N} \equiv |\vec f_{ij}\cdot
\hat r_{ij}|$, the tangential force: $f_{ij}^{\ T} \equiv |\vec
f_{ij} - (\vec f_{ij}\cdot \hat r_{ij})\hat r_{ij}|$. All
quantities are assumed properly a-dimensional for simplicity of
notation. The terms inside the brackets $\{\cdot\}$ correspond to
the jamming constraint function $\Theta_{\rm jam}$ in Eq.
(\ref{omega}), and therefore define the ensemble of jammed states.
The first three $\delta-$functions inside the big brackets impose
Newton's second and third law. The Heaviside $\Theta-$function
imposes the Coulomb condition and the last $\delta-$function
imposes the touching grain condition for hard spheres, assuming
identical grains of unit diameter. Integration is over all forces
and positions which are assumed to be equally probable as in the
flat average assumption of the micro-canonical ensemble.

The conditions specified in Eq. (\ref{exact}) are met in the
numerical packings, thus the results of Figs. \ref{numerics}b-d
can be interpreted as the ensemble average of Eq. (\ref{exact})
under assumption of uniformity in the jammed configurations.
However, Eq. (\ref{exact}) is difficult to solve. Analytical
progress can be done by considering a coarse-graining of the
Voronoi volume function and working with mesoscopic theory
\cite{jamming2} to obtain a configurational entropy at the
mesoscopic level.

{\it Coordination number Z-ensemble.---} Simple counting
arguments, neglecting correlations between nearest neighbors,
consider that a necessary condition for mechanical equilibrium is
that the number of independent force variables must be larger or
equal than the number of linear independent force/torque balance
equations.  Alexander \cite{alexander} conjectured that at the
transition point for frictionless spherical packings
\cite{alexander,stealing} the system is exactly isostatic with a
minimal coordination, $Z = 2d=6$ in 3d.  Such a conjecture can be
extended to the infinite friction case, where $Z = d+1=4$
\cite{stealing}. In the presence of finite inter-particle friction
coefficient $\mu$, the analytic form of $Z(\mu)$ is difficult to
achieve since the counting argument involves nonlinear inequality
constraints through the Coulomb condition. Despite the theoretical
difficulty, there exists a dependency of $Z$ with $\mu$ suggested
by simulations \cite{kertesz,jamming2,silbert}.

We consider a mesoscopic free volume function coarse-grained over
a few coordination shells that reduces the degrees of freedom to
the coordination number $z'$ as shown in \cite{jamming2}:
$w(z')=\frac{2\sqrt{3}}{z'} V_g$. The $w(z')$ function is
calculated \cite{jamming2} by obtaining the probability
distribution of Voronoi volumes associated with a particle in the
jammed state. This probability is decomposed into a bulk term
depending on $w$ and a contact term depending on $z'$. Note that
$z'$ refers to geometrical contacts that may carry no forces and
can be larger than the mechanical coordination $Z$ given by the
isostatic condition: $Z<z'$ \cite{jamming2}.

The mesoscopic approximation refers to the use of such a
mesoscopic free volume function in the partition function, taking
into account the effect of the environment of a particle rather
than the particle itself. Thus, we can reduce the partition
function to a single non-interacting particle.  The mesoscopic
entropy density is obtained in the canonical ensemble
\cite{sirsam,jamming2}:
\begin{equation}
s_{\rm meso}= \langle w \rangle /\lambda X + \lambda \ln
\int_{Z(\mu)}^6 g(z') \exp\left(-\frac{w(z')}{\lambda X}\right)
dz', \label{meso-entropy}
\end{equation}
where the isostatic condition enters through the limit of
integration. The space of configurations is considered discrete
since the states are collectively jammed \cite{salvatore1}. If the
typical distance between states is $h_z$, the density of states is
$g(z') \propto h_{z}^{z'}$ analogous to discretization of phase
space imposed by the Heisenberg principle with density of states
$h^{-d}$, where $h$ is the Planck constant. Equation.
(\ref{meso-entropy}) is a coarse-grained mesoscopic form of the
full entropy in Eq. (\ref{exact}) reducing the degrees of freedom
to $z$. Thus the effective "Plank constant" $h_z$ appears in Eq.
(\ref{meso-entropy}) and not in Eq. (\ref{exact}).

The mesoscopic entropy of Eq. (\ref{meso-entropy}) is plotted in
Fig. \ref{numerics}e as a function of $\phi$ for different values
of $Z$.  We see that it captures the general behavior found in the
simulations, i.e., maximal at RLP for $Z=4$ and minimal at RCP.
Furthermore, all the curves for different $Z$ approach $S\sim \ln
X$ as $X\to 0$, similar to a thermal ideal gas.  We conclude that,
at the mesoscopic level, the entropy vanishes at RCP (in fact it
diverges to $-\infty$ when $\phi\to \phi_{\rm RCP}$ closer than a
constant proportional to $h_z$), providing a characterization of
RCP.  This result qualitatively resembles behavior of the
complexity of the jammed state in the replica approach to jamming
\cite{parisi-kurchan}. We use $h_{z} = 0.01$ in Fig
\ref{numerics}e such that the mesoscopic entropy vanishes very
close to the predicted value of $\phi_{\rm RCP} \approx 0.634$
\cite{jamming2} and the value of the microscopic entropy as
calculated via graph theoretical methods.  The value of $h_z$ is
chosen to fit the theory with simulation as close as possible,
where the only constraint imposed by theory is $h_z < 1$. An
important result is the direct implication of a larger number of
states available to jammed systems at RLP with respect to any
higher volume fraction.

It is possible to interpret the RCP as a Kauzmann point (K-point)
\cite{parisi-kurchan} in analogy with the density (temperature) at
which the configurational entropy of a colloidal (molecular) glass
vanishes at the ideal glass transition.

We augment the mesoscopic entropy with the entropy of the
microscopic states to obtain the full entropy as $s =
s_\mathrm{meso} + s_{\rm micro}$. Since for frictionless packings
$s_\mathrm{meso}$ vanishes, then $s = s_\mathrm{micro}$, implying
that we can obtain $s_{\rm micro}$ from the full entropy of the
K-point as calculated numerically. Therefore, Fig. \ref{numerics}e
is plotted as $s_{\rm meso}/\lambda + 1.1 $ to obtain a plot of
the full entropy to compare with the full ensemble Eq.
(\ref{exact}) and simulations. This constant is obtained from the
value given by the Shannon Entropy, regarded as the entropy of
RCP, which is $1.1\lambda$.

In summary, a notion of jamming is presented that applies to
frictional hard spheres, as well as frictionless ones.  The entropy
reveals interesting features of the RCP and RLP states such as the
fact that RLP is maximally random with respect to RCP and that both
limits can be seen in terms of the entropy and equation of state.  The
theoretical model captures the shape of the entropy and the equation
of state, but not the actual values at RLP. Future study will be
devoted to capture the volume fluctuations not only at the mesoscopic
level but also at the microscopic level neglected in the present
theory, which may lead to an understanding of the finite value of
entropy found at RCP in the numerics. The agreement between theory and
simulation is sufficient to indicate that the methods presented herein
are appropriate for evaluating the entropy of jammed matter allowing
the characterization of the state of randomness of RLP and RCP. We
note that an interesting recent work \cite{ciamarra} finds random very
loose packings with negative effective temperature in an energy
ensemble approach. Such states (analogous to negative $X$ in the
volume ensemble) are discussed elsewhere \cite{jamming2}. Our results
are in general agreement with those of \cite{aste}.

{\bf Acknowledgements}. We thank the financial support of the NSF and
DOE, Office of Basic Energy Sciences.

\end{document}